\documentclass[conference]{IEEEtran}
\IEEEoverridecommandlockouts
\usepackage{cite}
\usepackage{amsmath,amssymb,amsfonts}
\usepackage{algorithmic}
\usepackage{graphicx}
\usepackage{textcomp}
\usepackage{xcolor}
\def\BibTeX{{\rm B\kern-.05em{\sc i\kern-.025em b}\kern-.08em
    T\kern-.1667em\lower.7ex\hbox{E}\kern-.125emX}}

\usepackage{url}            
\usepackage{booktabs}       
\usepackage{nicefrac}       
\usepackage{microtype}      
\usepackage[space]{grffile} 
\usepackage{subcaption}

\begin{document}

\title{Denoising Auto-encoder with Recurrent Skip Connections and Residual Regression for Music Source Separation\\
}

\author{\IEEEauthorblockN{Jen-Yu Liu}
\IEEEauthorblockA{\textit{Research Center for IT Innovation} \\
\textit{Academia Sinica}\\
Taipei, Taiwan \\
ciaua@citi.sinica.edu.tw}
\and
\IEEEauthorblockN{Yi-Hsuan Yang}
\IEEEauthorblockA{\textit{Research Center for IT Innovation} \\
\textit{Academia Sinica}\\
Taipei, Taiwan \\
yang@citi.sinica.edu.tw}
}


\maketitle

\begin{abstract}
Convolutional neural networks with skip connections have shown good performance in music source separation. In this work, we propose a denoising Auto-encoder with Recurrent skip Connections (ARC). We use 1D convolution along the temporal axis of the time-frequency feature map in all layers of the fully-convolutional network. The use of 1D convolution makes it possible to apply recurrent layers to the intermediate outputs of the convolution layers. In addition, we also propose an enhancement network and a residual regression method to further improve the separation result. The recurrent skip connections, the enhancement module, and the residual regression all improve the separation quality. The ARC model with residual regression achieves 5.74 siganl-to-distoration ratio (SDR) in vocals with MUSDB in SiSEC 2018. We also evaluate the ARC model alone on the older dataset DSD100 (used in SiSEC 2016) and it achieves 5.91 SDR in vocals.
\end{abstract}


\section{Introduction}


Music source separation aims at separating music sources such as vocals, drums, strings, or accompaniment from the original song. It can facilitate tasks that require  clean sound sources, such as music remixing and karaoke \cite{Rafii2018}. In this work, we introduce a new model that uses denoising auto-encoder with symmetric skip connections for music source separation.  Symmetric skip connections have been used for biomedical image segmentation \cite{Ronneberger2015} and singing voice separation \cite{Jansson2017}. Our model is different in that it uses 1D convolutions instead of 2D convolutions. Using 1D convolutions has the benefit that we can use recurrent layers right after the convolution layers. Furthermore, an enhancement module and a residual regression method are introduced in addition to the separation module.


\section{Proposed models} \label{sec:method}
In this section, we introduce the separation model, the enhancement model, and  residual regression. 

\subsection{Separation model} \label{sec:sep}

The separation model is a fully-convolutional network (FCN) \cite{Oquab2015,Long2015}. All the convolution layers use 1D convolution. We call it the ARC model, for it is in principal a denoising \underline{a}uto-encoder with \underline{r}ecurrent skip \underline{c}onnections.

\begin{figure}
	\centering
	\includegraphics[width=\columnwidth]{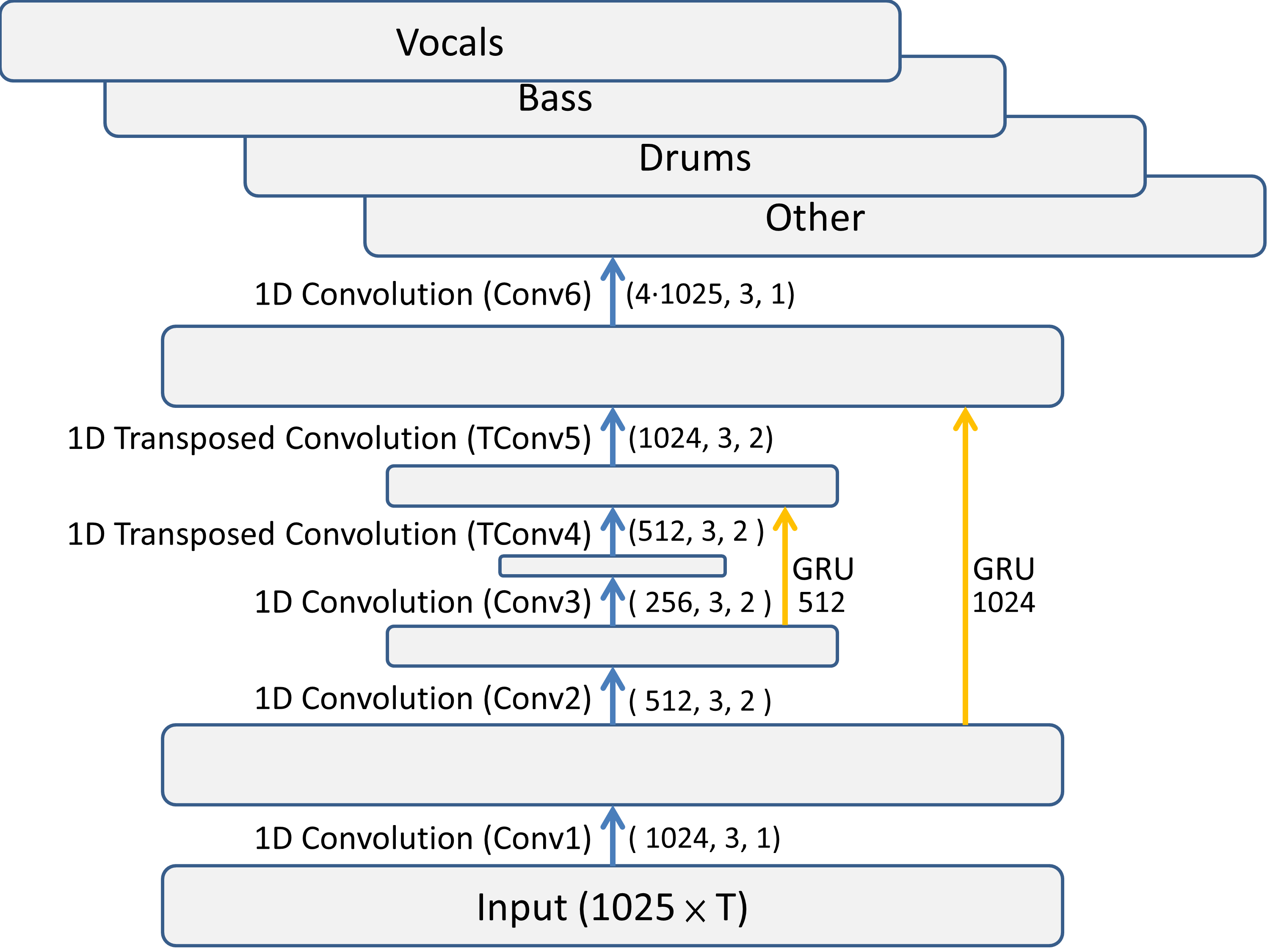}
	
	\caption{Diagram of the proposed separation model, ARC. Each tuple in the figure represents (output channels, filter size, stride) of the corresponding convolution layer. An STFT window size 2,048 is used and a spectrogram is symmetric in the frequency dimension, so the effective dimension of a spectrogram is $(2,048/2+1)=1025$. $T$ is the number of temporal frames.}
	\label{fig:ARC}
\end{figure}

CNN with symmetric skip connections had been used for singing voice separation by Jansson \emph{et al.} \cite{Jansson2017}. They used 2D convolutions in their convolutional neural networks (CNNs). The output tensor of a 2D convolution layer is of the shape (channels, frequency bins, temporal points). If we want to apply recurrent layers to this tensor, the dimension of frequency bins will pose some problems. 

In our model, the convolution layers use 1D convolutions, namely doing convolutions along the temporal axis \cite{Liu2016, Chou2018}. The output tensor of an 1D convolution layer takes the shape (channels, temporal points). This allows us to directly apply recurrent layers to the convolution output tensors.

The proposed architecture is presented in Fig. \ref{fig:ARC}. It contains six convolution layers and two skip connections. The two skip connections are processed by gated recurrent unit (GRU) layers \cite{Cho2014, Chung2014}. We use weight normalization \cite{Salimans2016} instead of batch normalization \cite{Ioffe2015} in each convolution layer. Leaky rectified linear units (Leaky ReLUs) with 0.01 slope \cite{Maas2013} are applied to all the convolution and transposed convolution layers.

The model takes a spectrogram of a song clip as the input. An input is also referred to as a mixture because it contains the sources such as vocals, bass, drums, and other sounds. The input to the model is $log(1+\text{mixture\_spectrogram})$, and the training target is the source spectrograms, that is, the concatenation of $log(1+\text{source\_1}), log(1+\text{source\_2}), ..., log(1+\text{source\_S})$, where $S$ denotes the number of sources. This model can be seen as a denoising auto-encoder because, for one target source, the other sources can be seen as noises in the mixture signal.

In our pilot experiments, we also tried to apply a softmax function to the output layer so that the network predicts masks for different sources and enforces the condition that the summation of the predicted source spectrograms is equal to the mixture spectrogram. We found that this setting largely speeds up the training process, but the result becomes much worse. Therefore, we decided to use a leakly ReLU as the nonlinearity function to the output layer to directly estimate the source spectrograms.

%

\begin{figure}
	\centering
	\includegraphics[width=\columnwidth]{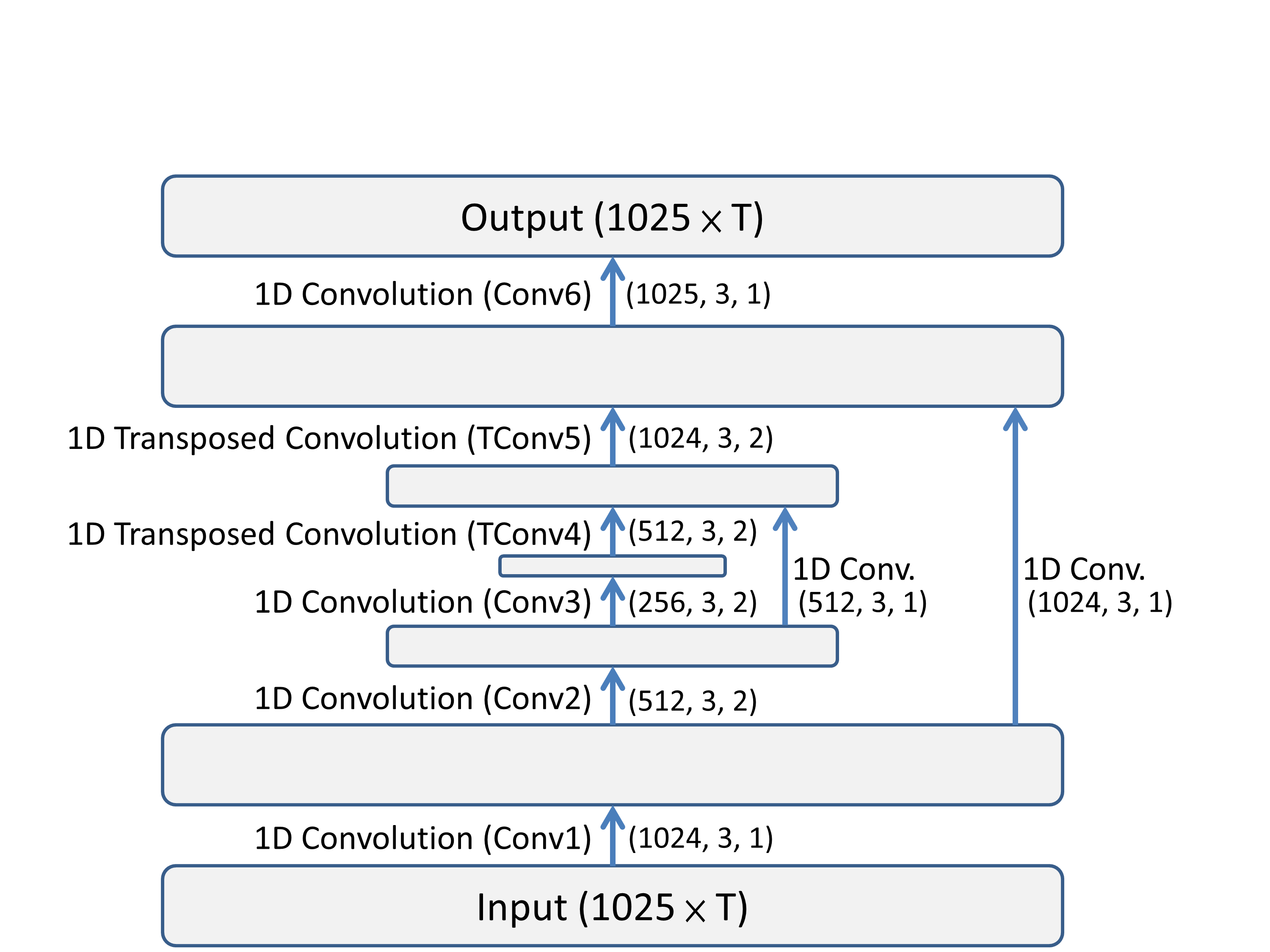}
	
	\caption{Diagram of the proposed enhancement model. Each tuple in the figure represents (output channels, filter size, stride) of the corresponding convolution layer.}
	\label{fig:enhancement}
\end{figure}

\begin{figure}
	\centering
	\includegraphics[width=\columnwidth]{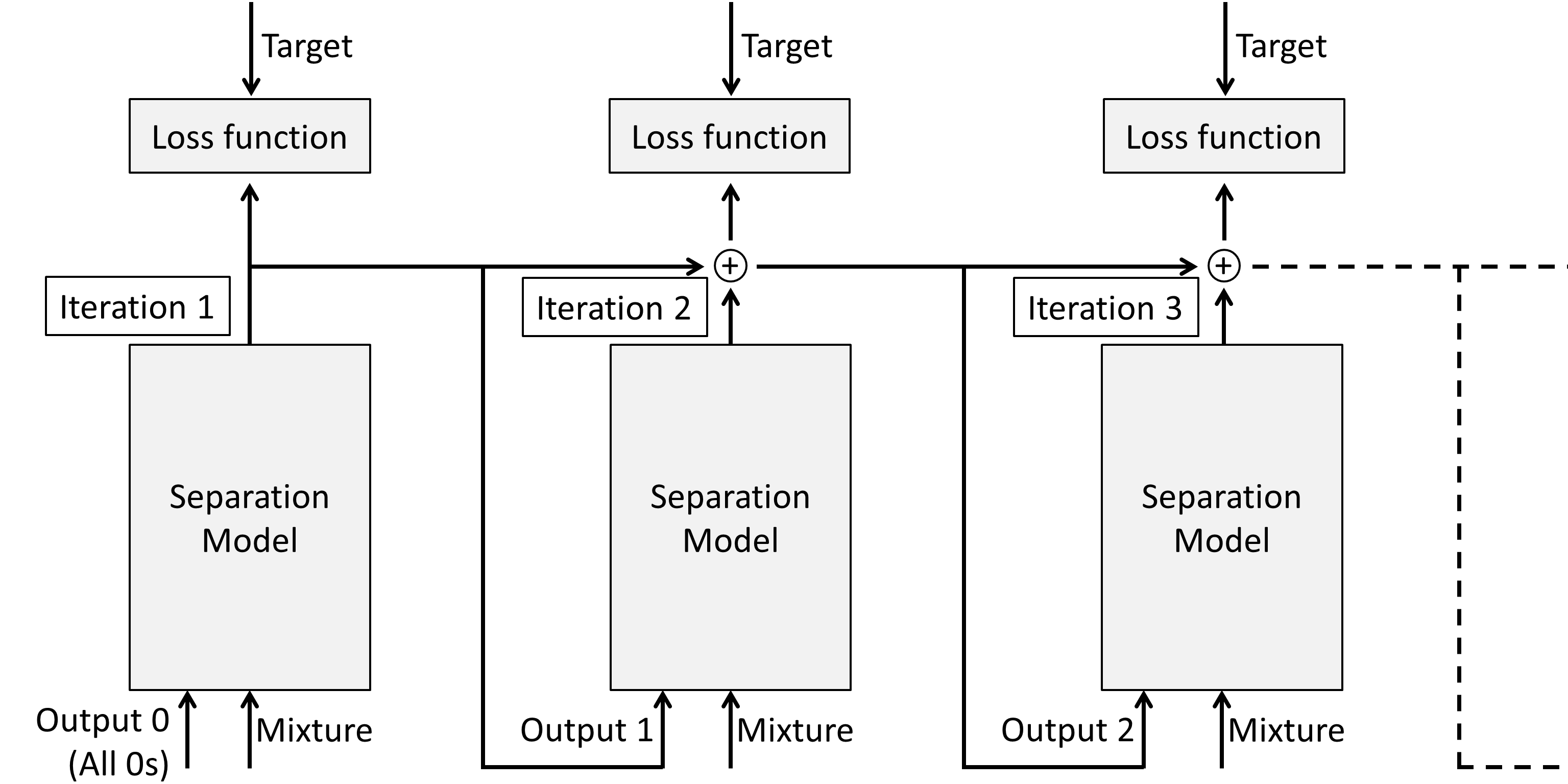}
	\caption{Illustration of residual regression. In iteration $i$, the separation model also takes the output $i-1$ as the input. The total output of iteration $i$ is the sum of the total output of iteration $i-1$ and the output of the separation model. Therefore, the separation model only has to estimate the residual.}
	\label{fig:rere}
\end{figure}

\subsection{Enhancement model}
The separation model is in charge of the task of music source separation. The small noises could be ignored in the training process because the losses introduced by other sources could be much larger than the losses introduced by the smaller artifacts. But, we human beings are very sensitive to those smaller artifacts, especially in vocals. 

In order to reduce these small artifacts, we introduce an extra enhancement model as a post-processing module. The enhancement model is another denoising auto-encoder that takes the output of a separation model (i.e. the ARC) as its input, and estimates an enhanced version of the separation result. Each source has its own enhancement model, and the training target is that specific source spectrogram.

The architecture is shown in Fig. \ref{fig:enhancement}. It is similar to ARC but the skip connections are implemented as convolution layers for simplicity. In the training process of the enhancement model, the parameters of a separation model are fixed.

\subsection{Residual regression}

Residual regression is also used to improve the separation result. Unlike the enhancement model, the model with residual regression uses the separation model itself to further improve the separation result.

The process of residual regression is depicted in Fig. \ref{fig:rere}. The separation model in Fig. \ref{fig:rere} is similar to the one introduced in Section \ref{sec:sep}. The difference is that the separation model takes another input feature map (the left arrow below the separation model) that is the output from the previous iteration. In iteration $i$, the separation model takes both the output $i-1$ and the mixture feature map as the input. For the iteration 1, the output 0 is set to an all-zero tensor with the same shape as the mixture feature map. The total output of iteration $i$ is the output of the separation model plus the total output of iteration $i-1$. In this way, the separation model will only estimate the residual of the target sources. In the training process, the total loss is the average of the losses from all the iterations.

\section{Evaluation}

The evaluation is conducted by using the official dataset MUSDB (100 songs for training and 50 songs for testing) and the official packages\footnote{\url{https://github.com/sigsep/sigsep-mus-eval} and \url{https://github.com/sigsep/sigsep-mus-2018-analysis}} from SiSEC2018 \cite{sisec2018}. The models are implemented with PyTorch.\footnote{\url{https://pytorch.org/}} We will report the evaluation result in terms of signal-to-distortion ratio (SDR) \cite{Nugraha2016}, as it is the most widely used metric in literature \cite{Uhlich2017,Liutkus2017,sisec2018}

\begin{table*}
	\caption{Performance (in SDR) for MUSDB in SiSEC 2018}
	\label{tab:submissions}
	\centering
	\begin{tabular}{llll|ccccc}
		
		SiSEC ID	& Skip connections	& Enhancement	& Residual fegression	& vocals	& drums	& bass	& other	& accompaniment \\
		\hline
		JY1     	& 1 GRU layer		& No			& No	& 5.57  	& 4.60  & 3.18  & 3.45 	& 11.81 \\
		JY2     	& 1 GRU layer		& Yes			& No	& 5.69  	& 4.76  & 3.58  & 3.70  & 11.90 \\
		JY3			& 1 GRU layer		& No			& Yes (3 iterations)& 5.74		& 4.66  & 3.67  & 3.40  & 12.08 \\
	\end{tabular}
\end{table*}

\subsection{Training process}
The training dataset is MUSDB.\footnote{\url{https://sigsep.github.io/datasets/musdb.html#tools}} It contains 100 songs, each of which has four sources: drums, bass, other, and vocals. We randomly choose 90 songs as the training set and 10 songs as the validation set. The validation set is used for early stopping. Each song is divided into 5-second sub-clips.

The short-time Fourier transform (STFT) is applied to the sub-clips for feature extraction. The native sampling rate 44,100 is used with a window size 2,048 and a hop size 1,024.

Uhlich \emph{et al.} \cite{Uhlich2017} showed that data augmentation is crucial to compensate for the scarcity of training data in music source separation. We conduct the online data augmentation to increase the number of training data as follows. Assume we have $N$ 5-second sub-clips. First, we randomly choose one sub-clip from the $N$ sub-clips for each source. Note that the sub-clip chosen for one source could be different from the sub-clip chosen for another source. The four sub-clips from the four sources are summed, leading to the mixture of one training instance. Then, we use the spectrogram of this mixture as the input and use the concatenated spectrograms of the four source sub-clips as the training target.

We use mean square error (MSE) as the loss function for updating the network. Assume that the mini-batch size is $B$, and there are $S$ sources, $T$ temporal points, and $F$ frequency bins. Then, the loss function is $(\sum_{b=1}^B \sum_{s=1}^S \sum_{t=1}^T \sum_{f=1}^F|P_{b,s,t,f}-log(1+G_{b,s,t,f})|^2)/(BSTF)$, where $P_{b,s,t,f}$ is the prediction and $G_{b,s,t,f}$ is the target source spectrogram.


We use Adam \cite{Kingma2014} and a mini-batch of 10 instances to train the models. The initial learning rate is set to 0.001 for the convolution layers, and it is set to 0.0001 for the GRU layers. We found that using 0.001 learning rate often lead to gradient explosion for the GRU layers, while the training process was stable when we used 0.0001 for the GRU layers. 

\subsection{Testing process}

In the testing phase, an entire song is processed at once. Because we adopt a FCN design, our model can deal with songs of abitrary length. Multi-channel Wiener filter is used for post-processing \cite{Nugraha2016, Uhlich2017}. We use the phases of the mixture to convert the estimated source spectrograms into waveforms via the inverse STFT. We use the sum of the estimates of the four sources as the estimate of the accompaniment (`accomp.').

\subsection{Result}

In this subsection, we show the performance of our submissions to SiSEC2018. The result is shown in TABLE \ref{tab:submissions}. In the model with residual regression (JY3), we run three iterations. We can see from this table that JY2 (using enhancement model) and JY3 (using residual regression) improves over JY1 in almost all sources.

\begin{figure}
	\centering
	\includegraphics[width=0.8\columnwidth]{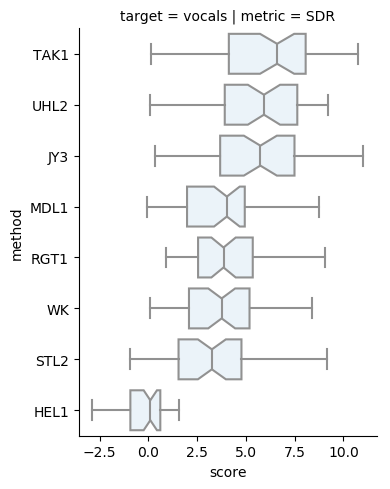}
	\caption{Result (in SDR for vocals) for submissions of SiSEC 2018. This figure shows the best supervised model from each author group without using additional training data. }
	\label{fig:sisec}
\end{figure}

Fig. \ref{fig:sisec} display the SiSEC 2018 results of the models using supervised approaches without using additional training data, showing the best model of each author group.\footnote{This figure is generated with a modified version of the code provided by the organizers \url{https://github.com/sigsep/sigsep-mus-2018-analysis}. We specify ``not using additional training data'' here, because some submissions did use additional training data (not by data augmentation but by actually including more songs with clean sources for training.}  Statistically the result of JY3 in vocals is not significantly different from that of the other two leading models TAK1\footnote{\url{https://github.com/sigsep/sigsep-mus-2018/blob/master/submissions/TAK1/description.md}} \cite{Takahashi2017,Takahashi2018} and UHL2\footnote{\url{https://github.com/sigsep/sigsep-mus-2018/blob/master/submissions/UHL2/description.md}} \cite{Uhlich2017}, according to the official SiSEC2018 report \cite{sisec2018}.

\subsection{Effect of different skip connections}

\begin{figure*}
	\centering
	\begin{subfigure}{0.5\textwidth}
		\centering
		\includegraphics[width=\columnwidth]{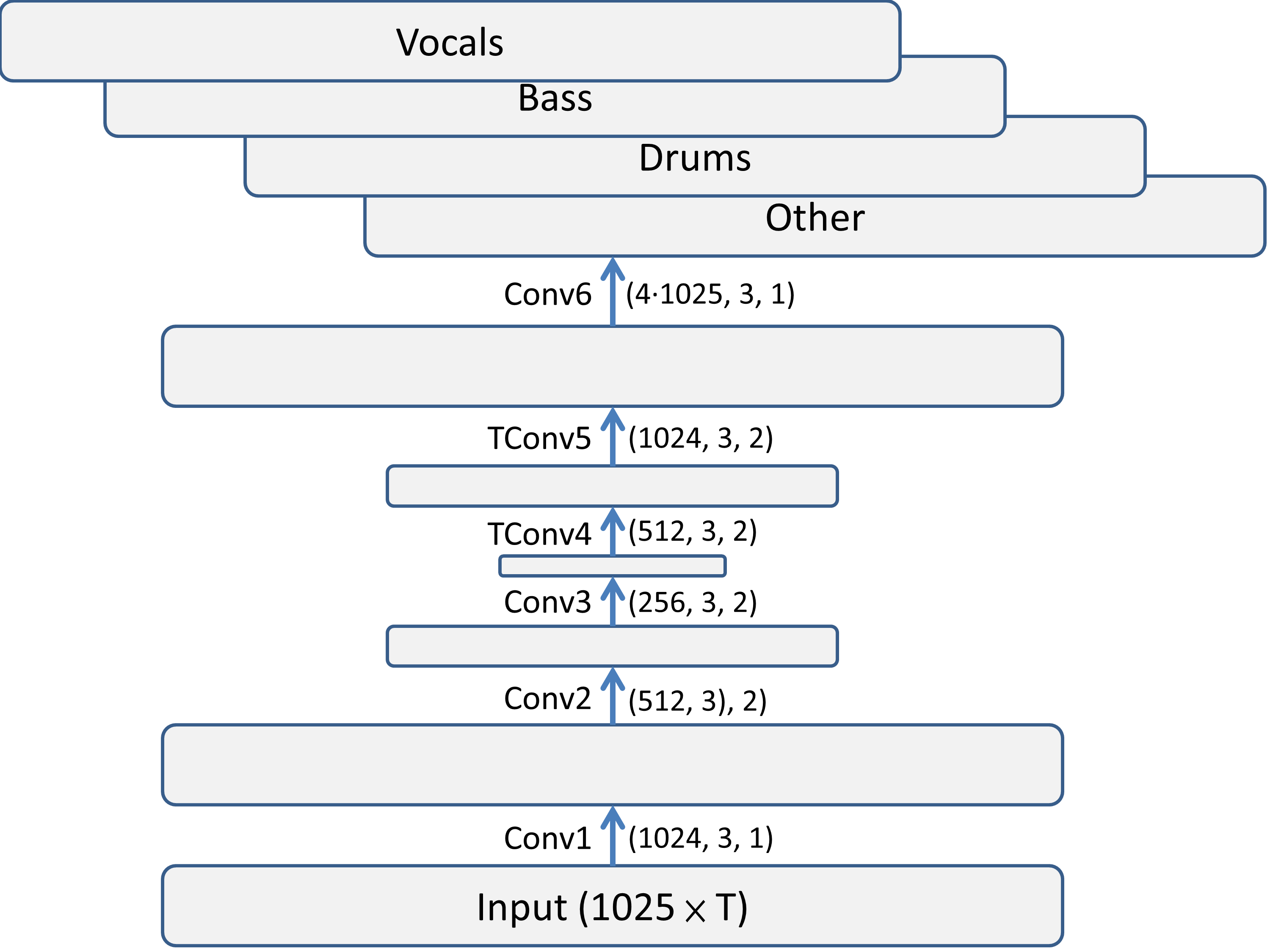}
		\caption{No skip connections}
	\end{subfigure}%
	\begin{subfigure}{0.5\textwidth}
		\centering
		\includegraphics[width=\columnwidth]{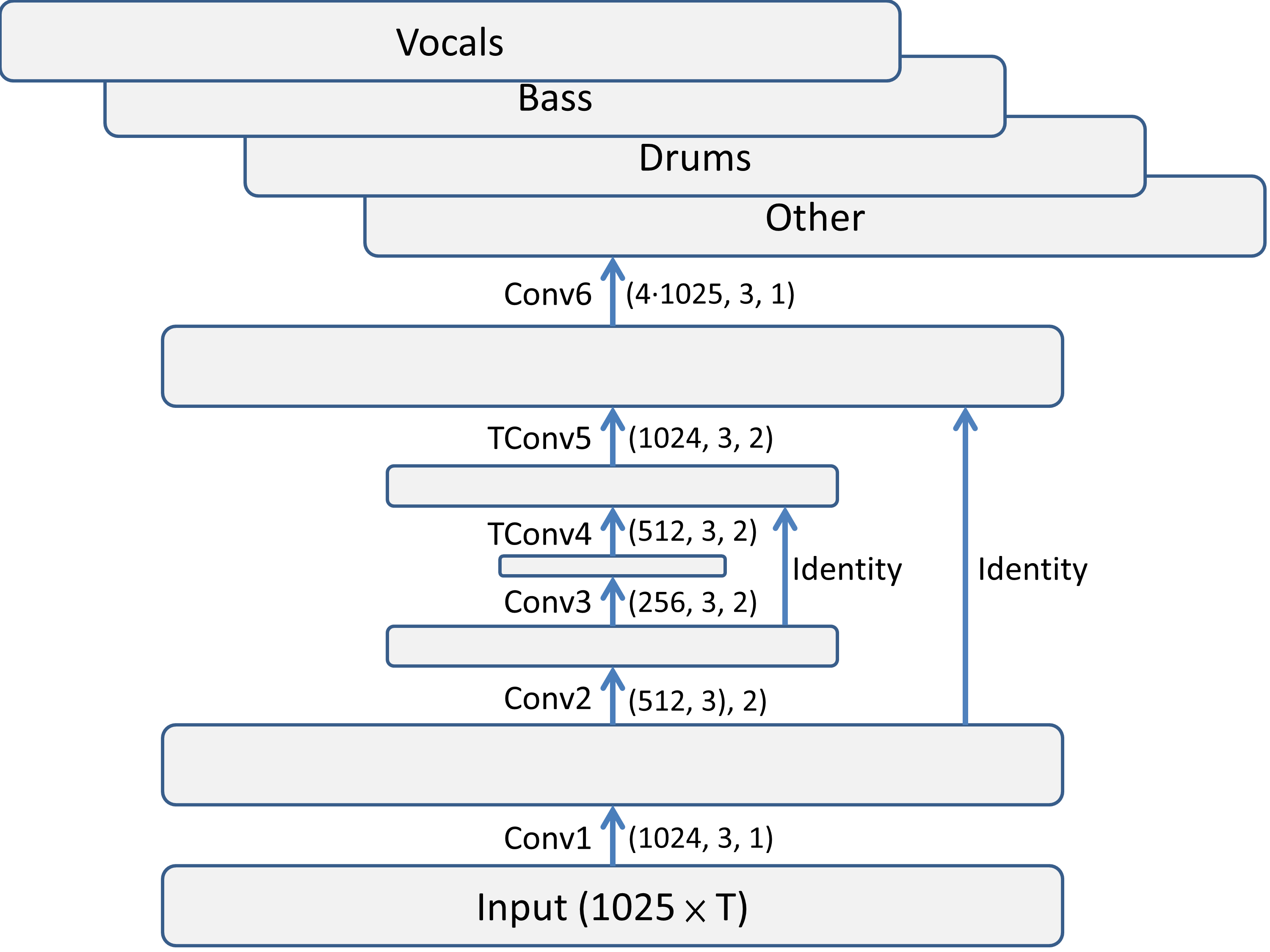}
		\caption{Identity skip connections}
	\end{subfigure}%
	\par\bigskip
	\begin{subfigure}{0.5\textwidth}
		\centering
		\includegraphics[width=\columnwidth]{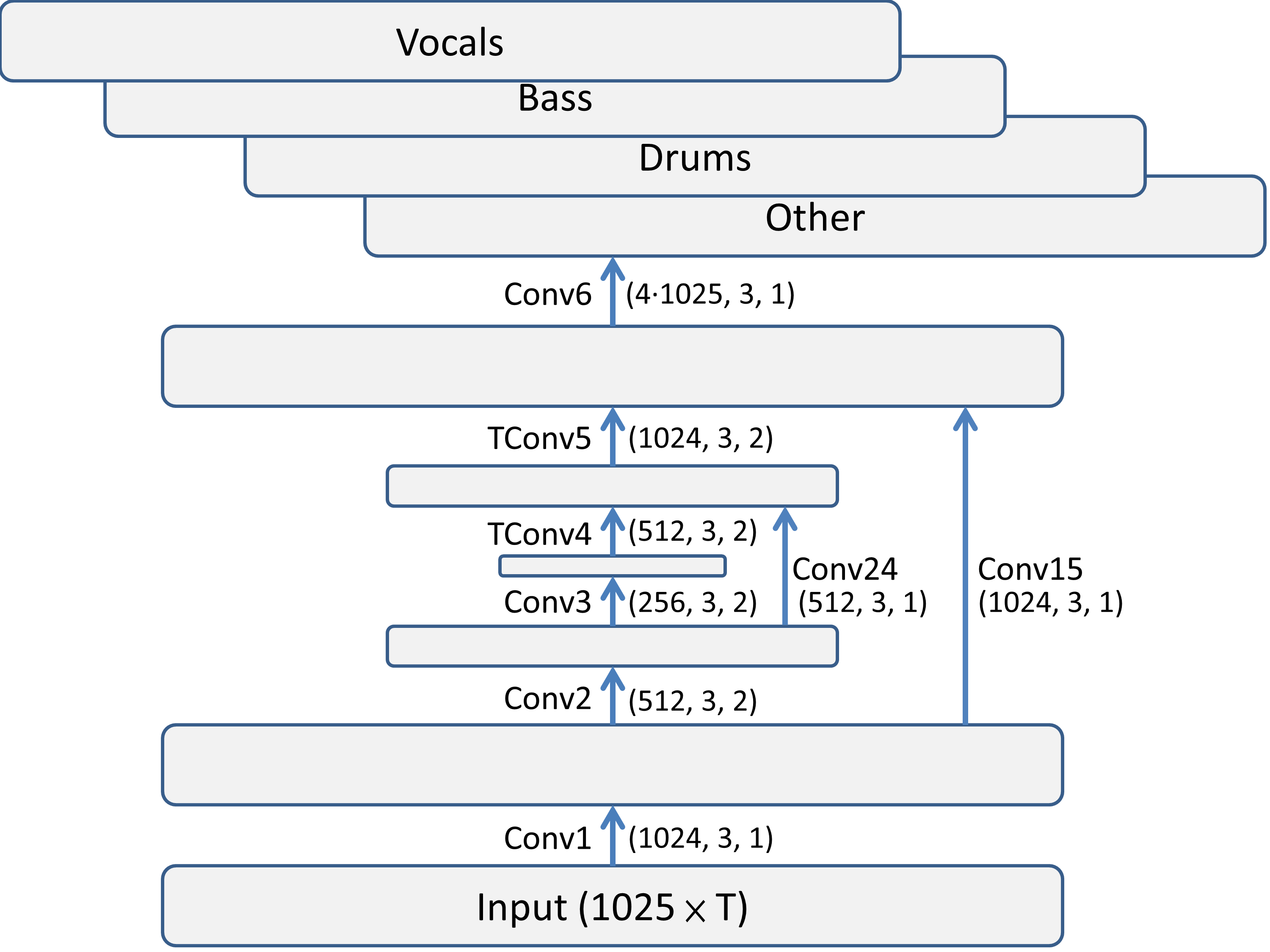}
		\caption{Convolution skip connections}
	\end{subfigure}%
	\begin{subfigure}{0.5\textwidth}
		\centering
		\includegraphics[width=\columnwidth]{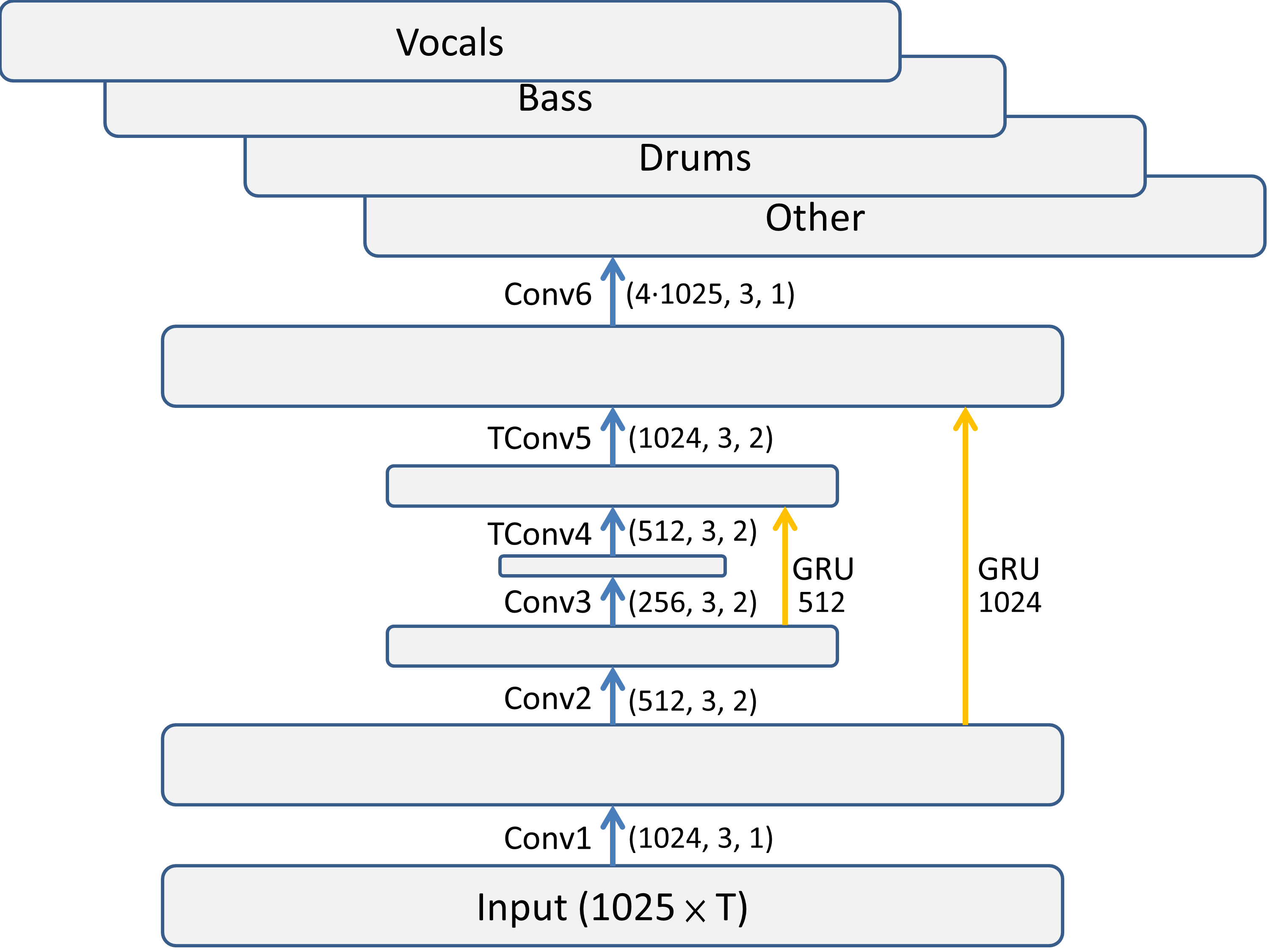}
		\caption{Recurrent skip connections (ARC)}
	\end{subfigure}%
	
	\caption{Different skip connections}
	\label{fig:skip}
\end{figure*}

We compare different skip connections in this subsection. The four compared architectures are shown in Fig. \ref{fig:skip}, and the result is shown in TABLE \ref{tab:skip}. 
We can see that the models with skip connections outperform the one without skip connections, and the model with recurrent skip connections outperforms the one with convolution skip connections.

\begin{table}
	\caption{Comparison of different skip connections (in SDR) for MUSDB in SiSEC 2018}
	\label{tab:skip}
	\centering
	\begin{tabular}{l|ccccc}
		Skip connections	& vocals	& drums	& bass	& other	& accomp. \\
		\hline
		None				& 4.41  	& 4.48  & 3.43  & 2.91  & 10.74 \\
		Direct (identity)	& 5.05  	& 4.65  & 3.41  & 3.02  & 11.25 \\
		1 Convolution layer	& 5.03  	& 4.78  & 3.37  & 2.80  & 11.39 \\
		1 GRU layer (JY1)	& 5.57  	& 4.60  & 3.18  & 3.45 	& 11.81 \\
	\end{tabular}
\end{table}

\subsection{Applying recurrent layers at different locations}

\begin{figure*}
	\centering
	\begin{subfigure}{0.45\textwidth}
		\centering
		\includegraphics[width=\columnwidth]{img/ARC_simple.png}
		\caption{Recurrent skip connections (ARC)}
	\end{subfigure}%
	\begin{subfigure}{0.45\textwidth}
		\centering
		\includegraphics[width=\columnwidth]{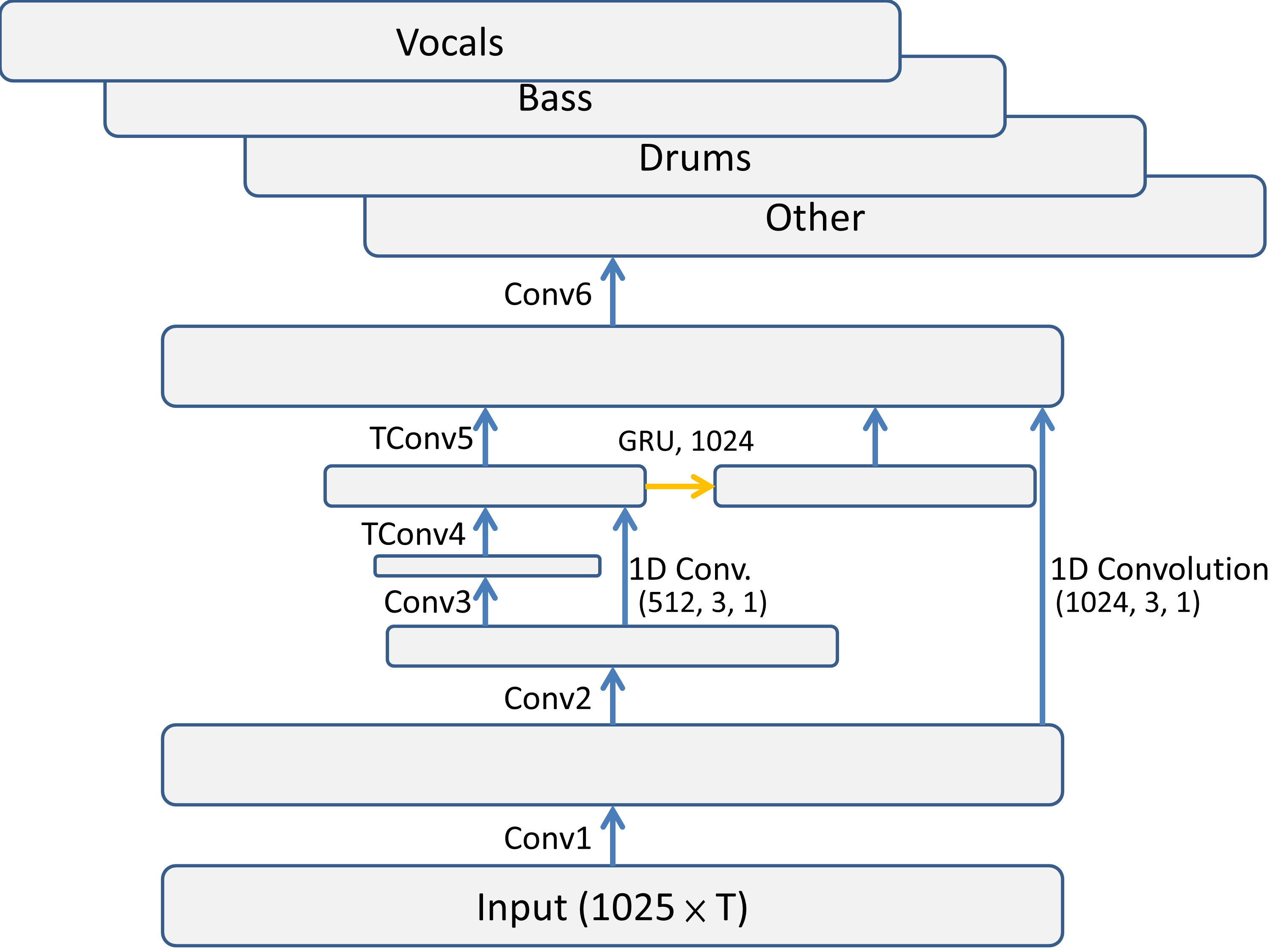}
		\caption{Pre-output recurrent layer}
		\label{fig:preoutput_recurrent}
	\end{subfigure}
	
	\caption{Recurrent layers at different locations. The yellow arrows indicate recurrent layers.}
	\label{fig:recurrent}
\end{figure*}

The recurrent layers could be applied at different locations of the separation model. We tested several possibilities, and many of them improves over the non-recurrent versions. For example, another possible way of using recurrent layers is shown in Fig. \ref{fig:preoutput_recurrent} and its performance is shown in TABLE \ref{tab:recurrent}. Among these  variants, we found that applying the recurrent layers to the skip connections is the most effective one.

\begin{table}
	\caption{Recurrence at different layers (in SDR) for MUSDB in SiSEC 2018}
	\label{tab:recurrent}
	\centering
	\begin{tabular}{l|ccccc}
		Where to use 					& 			&		&		&		&	\\
		recurrent layers				& vocals	& drums	& bass	& other	& accomp. \\
		\hline
		Skip connections (JY1)			& 5.57  	& 4.60  & 3.18  & 3.45 	& 11.81 \\
		After TConv4 output				& 5.36  	& 4.38  & 3.53  & 3.66  & 11.91 \\
		
	\end{tabular}
\end{table}

\subsection{Batch normalization VS Weight normalization}\label{sec:norm}

We have found that the separated audios subjectively sound less noisy using weight normalization \cite{Salimans2016} in convolution layers than the separated audios using batch normalization \cite{Ioffe2015} after convolution layers. However, the objective evaluation with SDR suggests that their results are very close in vocals and the one with batch normalization is even better in the other sources, as shown in TABLE \ref{tab:norm}. 

\begin{table}
	\caption{Batch normalization VS Weight normalization  (in SDR) for MUSDB in SiSEC 2018}
	\label{tab:norm}
	\centering
	\begin{tabular}{l|ccccc}
		
		Normalization		& vocals	& drums	& bass	& other	& accomp. \\
		\hline
		Weight norm (JY1)   & 5.57  	& 4.60  & 3.18  & 3.45 	& 11.81 \\
		Batch norm			& 5.56  	& 4.92  & 3.63  & 3.57  & 11.98 \\
	\end{tabular}
\end{table}

\subsection{Qualitative Result}

\begin{figure*}
	\centering
	\begin{subfigure}{0.95\textwidth}
		\centering
		\includegraphics[width=\textwidth]{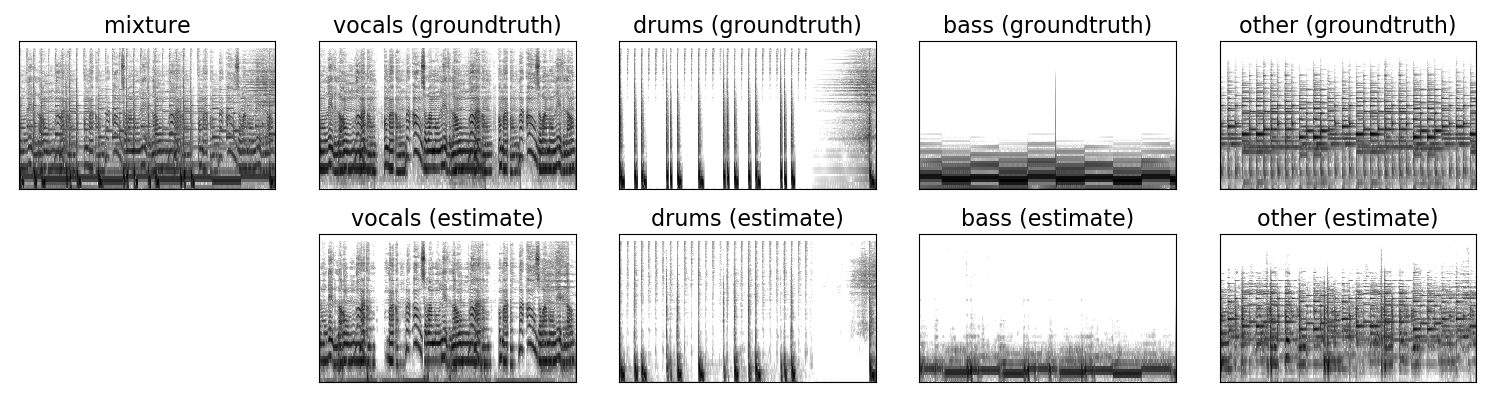}
		\caption{Little Chicago's Finest -- My Own from MUSDB test set, 30 seconds to 40 seconds}
	\end{subfigure}%
	\par\bigskip
	\begin{subfigure}{0.95\textwidth}
		\centering
		\includegraphics[width=\textwidth]{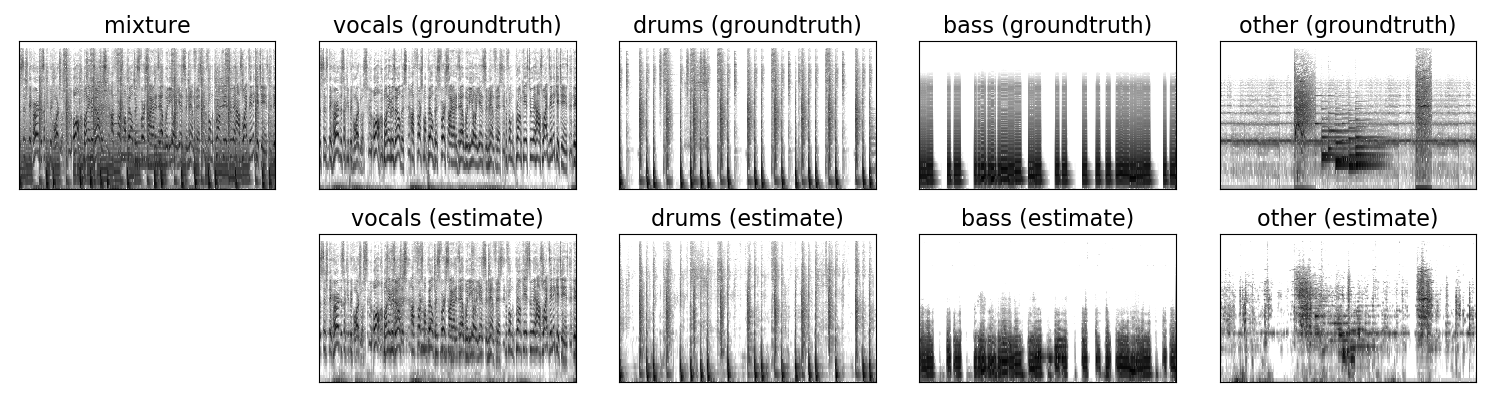}
		\caption{Side Effects Project -- Sing With Me from MUSDB test set, 30 seconds to 40 seconds}
	\end{subfigure}%

	\caption{Examples of spectrograms of the groundtruth sources and the estimated sources for two songs in the test set of MUSDB used by SiSEC 2018. The first row contains the groundtruth sources and the second row contains the estimated sources by the model with residual regression (JY3). The first column shows the original song, that is, the mixture. }
	\label{fig:spectrograms}
\end{figure*}

Fig. \ref{fig:spectrograms} shows the groundtruth spectrograms and the estimated spectrograms of two example songs from the MUSDB test set. The groundtruths and the estimates have similar patterns. We can see clear activations of the fundamental frequencies and their harmonics from the estimated spectrograms. On the other hand, we can observe that the estimated spectrograms are less sharp and noisier compared to the groundtruth spectrograms, which indicate rooms for improvement in the future work.

We also build a website (\url{http://mss.ciaua.com}) to demo the result of the proposed model JY3 for songs not in MUSDB.

\subsection{Evaluating with DSD100 dataset}
We also evaluate the proposed ARC net with DSD100 dataset that was used in SiSEC2016 \cite{Liutkus2017}. We evaluate ARC with batch normalization as introduced in Section \ref{sec:norm} with DSD100 by using the official toolkit.\footnote{\url{https://github.com/faroit/sisec-mus-results}} 
The enhancement and residual regression are not used in this evaluation. 
We use the 50/50 train/test split specified by SiSEC2016. The result is shown in TABLE \ref{tab:dsd100}. The result of our model is only second to that of the MMDenseNet \cite{Takahashi2017} and MMDenseLSTM \cite{Takahashi2018} models proposed by Takahashi \emph{et al.} The TAK1 method shown in Fig. \ref{fig:sisec} is an extended version of these models.

\begin{table}
	\caption{Evaluation on DSD100 (in SDR). We use ARC with batch normalization for our model here.}
	\label{tab:dsd100}
	\centering
	\begin{tabular}{l|ccccc}
	    & vocals& drums	& bass	& other	& accomp. \\
		\hline
		DeepNMF \cite{LeRoux2015}                   & 2.75	& 2.11	& 1.88	& 2.64	& 8.90 \\
		NUG \cite{Nugraha2016}						& 4.55	& 3.89	& 2.72	& 3.18	& 10.29 \\
		MaDTwinNet \cite{Drossos2018}               & 4.57	& ---	& ---	& ---	& --- \\
		BLSTM \cite{Uhlich2017}						& 4.86	& 4.00	& 2.89	& 3.24	& 11.26 \\
		SH-4stack \cite{Park2018}                   & 5.16	& 4.11	& 1.77	& 2.36	& 12.14	\\
		BLEND \cite{Uhlich2017}						& 5.23	& 4.13	& 2.98	& 3.52	& 11.70 \\
		MMDenseNet \cite{Takahashi2017}             & 6.00  & 5.37  & 3.91 & 3.81 & 12.10 \\
		MMDenseLSTM \cite{Takahashi2018}			& 6.31	& 5.46	& 3.73	& 4.33	& 12.73	\\
		Ours 		& 5.91  & 4.11  & 2.54  & 3.53 	& 11.31 \\
	\end{tabular}
\end{table}

\section{Conclusions} \label{sec:conclusion}
In this paper, we have presented our models for music source separation. We proposed to use 1D convolutions in convolution layers so that we can naturally apply recurrent layers to the convolution outputs. The experiments show that the recurrent skip connections largely improve the separation result. Moreover, the proposed enhancement model and residual regression can further improve the separation result.

For future work, we would be interested in applying the source separation models for other applications, such as singing style transfer \cite{Wu2018}, vocal melody extraction \cite{Bittner2017,Su2018}, instrument recognition \cite{Hung2018}, and lyrics transcription \cite{Tsai2018}.

\bibliographystyle{IEEEtran}
\bibliography{IEEEabrv,mss}

\end{document}